\documentclass{article}
\usepackage[backend=biber, 
    style=apa, uniquelist=false
  ]{biblatex}

\usepackage{hyperref}
\usepackage{cleveref} 
\usepackage{caption}
\usepackage{array}
\usepackage{titlesec}
\usepackage{multirow}
\usepackage{tikz}
\usepackage{xcolor}

\usepackage{pgfplots} 
\usepackage[inline]{enumitem}
\usepackage[letterpaper]{geometry}
\usepackage{hicss}
\usepackage{times}
\usepackage[none]{hyphenat}
\usepackage{url}
\usepackage{latexsym}
\usepackage{fancyhdr}

\newcommand{\etal}{\textit{et al}. }

\newcommand{\eg}{\textit{e}.\textit{g}. }

\definecolor{blue1}{RGB}{100,180,255}
\definecolor{red1}{RGB}{255, 60, 60}
\definecolor{green1}{RGB}{0, 193, 0}

\usetikzlibrary{patterns}

\fancypagestyle{firststyle}
{
\fancyhead{}

\fancyfoot[CE,CO]{\footnotesize M. Rebol et al., "Collaborative System Design of Mixed Reality Communication for Medical Training,"\\This paper has been accepted for the upcoming 56th Hawaii International Conference on System Sciences (HICSS-56). \href{https://hdl.handle.net/10125/102680}{https://hdl.handle.net/10125/102680}.}
\fancyfoot[LE,RO]{\thepage}
}

\title{Collaborative System Design of\\Mixed Reality Communication for Medical Training}

 \author{
  Manuel Rebol \\
 American University,\\Graz University of Technology \\
 {\underline{mrebol@american.edu}} \\  \\ 
 Adam Rutenberg\\
 George Washington\\University \\
 {\underline{ arutenberg@mfa.gwu.edu} } \\ \And
Krzysztof Pietroszek \\
 American University \\
 {\underline{pietrosz@american.edu}} \\ \\
\\Neal Sikka\\
 George Washington\\University \\
 {\underline{ nsikka@mfa.gwu.edu} } \\  \And
Claudia Ranniger\\
 George Washington\\University \\
 {\underline{ ranniger@gwu.edu} } \\ \\
Christian Gütl\\
 Graz University of Technology \\
 {\underline{c.guetl@tugraz.at} } \\ \And
 Colton Hood \\
 George Washington\\University \\
 {\underline{ chood@mfa.gwu.edu} } \\
}

\date{}

\begin{document}

\twocolumn[{%
\renewcommand\twocolumn[1][]{#1}%
\maketitle
\begin{center}
    \centering
    \captionsetup{type=figure}
    \includegraphics[width=\textwidth]{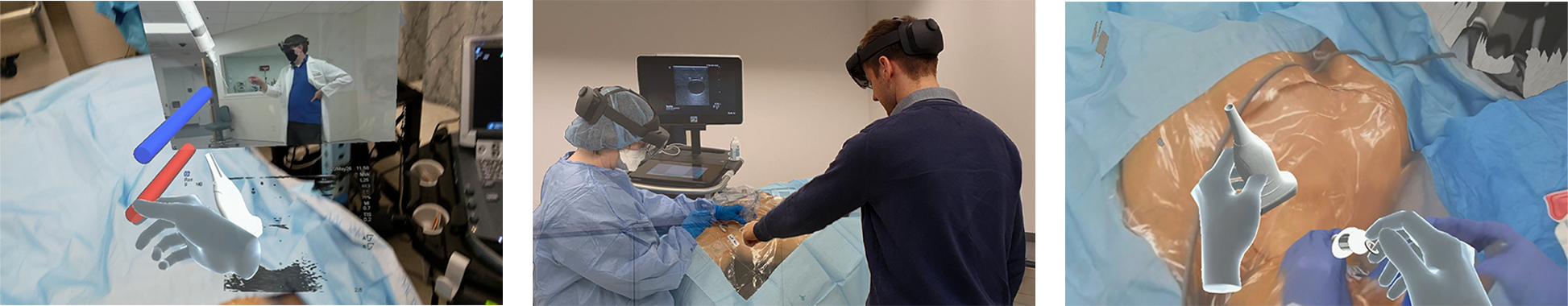}
    \captionof{figure}{Left: A local trainee performing a procedure guided by a virtual hand of the remote expert. Center: The remote expert visually interacting with the local operator. Right: The remote instructor's volumetric view of the local operators environment.\label{fig:teaser}}
\end{center}%
\vspace{0.5cm}
 }]
 \thispagestyle{firststyle}

\begin{abstract}
We present the design of a mixed reality (MR) telehealth training system that aims to close the gap between in-person and distance training and re-training for medical procedures. Our system uses real-time volumetric capture as a means for communicating and relating spatial information between the non-colocated trainee and instructor. The system's design is based on a requirements elicitation study performed in situ, at a medical school simulation training center. The focus is on the lightweight real-time transmission of volumetric data -  meaning the use of consumer hardware, easy and quick deployment, and low-demand computations. We evaluate the MR system design by analyzing the workload for the users during medical training. We compare in-person, video, and MR training workloads. The results indicate that the overall workload for central line placement training with MR does not increase significantly compared to video communication. Our work shows that, when designed strategically together with domain experts, an MR communication system can be used effectively for complex medical procedural training without increasing the overall workload for users significantly. Moreover, MR systems offer new opportunities for teaching due to spatial information, hand tracking, and augmented communication. 
\end{abstract}

\subsubsection*{Keywords:}

%
Mixed Reality, Augmented Reality, System Design, Telemedicine, Volumetric Communication

\section{Introduction}
Medical procedural skills are neither uniformly distributed within organizations nor around the world \autocite{Wachs2021}. We present the design process of a mixed reality (MR) communication system that facilitates remote procedural skill training. The objective is to make a step towards more equally distributed procedural skills among medical personnel, or access to an expert to bridge the gap. Through MR, training and emergency assistance can be given to remote geographic locations. Online MR communication offers an environment-friendly, inexpensive way to distribute procedural skills.  

While we focus on healthcare procedures, the issue of training and assistance by an expert who is not on-site is present across multiple disciplines and in many domains that depend on operator cognitive and manual procedural skills \autocite{hadar2011hybrid, rebol2021remote}. When mastery level skill must be brought to a remote location during natural disasters, epidemics, equipment breakdowns, etc., our system could be sent to the remote location or travel with a novice trainee to an area of need. 

In healthcare, the provision of in-person training for emergent situations in underfunded health systems is often not feasible \autocite{Wachs2021}. Few solutions have been proposed to date to resolve the gap between the quality and learning outcomes of distance training versus in-person training. The paucity of solutions presents a significant issue for equitable care delivery across health systems that have fewer experienced medical professionals \autocite{doi:10.34197/ats-scholar.2021-0053OC}.

When teaching procedural skills, it is important to simulate the setting to similar real world situations.
Combined with high-fidelity simulation equipment, MR and volumetric communication may offer a great opportunity for effective medical procedure training \autocite{article}, \autocite{s17102294}, \autocite{doi:10.1177/1553350620934931}.

In this paper, we focus on the design process of an MR system. Our main contribution is the strategic design process of an MR system for a specific task, the ultrasound-guided placement of a central venous catheter (US-CVC). We hypothesize that, when designed iteratively in consultation with domain experts, MR systems can be used intuitively by users not familiar with the technology and lower the workload during procedural training compared to video communication. System engineers need to understand the domain for which they are designing the system. Thus, we conduct an elicitation study (\Cref{sec:elicitation}) to understand the requirements of an MR system for US-CVC. The elicitation study analyzes in-person training with a specific focus on visual communication, mainly gestures. Moreover, the workload of instructors and trainees is measured during the procedure to determine a baseline. Based on the analysis from the elicitation study, we identify requirements ( \Cref{sec:sys-requirements}) for the MR system. The components and the implementation of the system based on the requirements are outlined in \Cref{sec:mr-system}. Finally, we evaluate the MR system in a user study (\Cref{sec:evaluation}) in which we compare it against video training. The workload is analysed using the NASA task load index (NASA TLX) \autocite{hart2006nasa} and the simulation task load index (SIM TLX) \autocite{Harris2020}, \autocite{10.1002/bjs.10795} instruments.

\section{Related Work} 
First, we compare different approaches toward augmented reality (AR), MR, and virtual reality (VR) system design. We find that the systems previously introduced do not meet the requirements that we identified in the elicitation study for the US-CVC procedure. Second, we analyze how similar systems were evaluated and discover a lack of uniformity. We use the validated instruments NASA TLX and SIM TLX for our system design process to compare medical procedural training through the different methodologies, in-person, videoconferencing-based, and MR teaching.

\subsection{Mixed Reality System Design}
The first remote procedural assistance systems using HMDs offered remote 2D video annotations that are locally spatially augmented \autocite{Chang2015, Lin2018}. Later, Zillner \etal \autocite{Zillner2018} suggested MR 3D assistance for remote repair assistance from a person on a remote computer that allows for spatial information for the remote expert. 
However, according to our system requirement analysis \Cref{sec:sys-requirements}, neither remote 2D assistance nor annotation-only assistance on a computer is sufficient for the US-CVC procedure.
Smartphone-based assistance designs \autocite{Lee2021, Gao2018} offer a low-cost and high accessibility solution. Yet, they are impractical because the local operator needs to use both hands during the US-CVC procedure.
The system by Lee \etal \autocite{Lee2021} shows promising evaluation results but does not send real-time updates to the remote operator which is crucial for emergency medical procedures.
The ARTEMIS system \autocite{artemis} uses multiple extended reality (XR) devices and, thus, supports high-accuracy medical assistance. 
A less complex system compared to ARTEMIS that shares components of our proposed design was introduced by Roth \etal \autocite{Roth2021}. 
Previous US-CVC studies have identified potential areas to augment the trainee's SBT. Chen \etal demonstrated that during US-CVC placement, experienced operators fixate on the ultrasound image for significantly more time than novices, who focus on tools \autocite{chen2018looks}. The ability of a mentor to independently view the procedural space is critical for patient safety.

\subsection{Workload Evaluation}
One of the most common limitations of previous studies is the lack of uniformity in evaluation methodology. Specifically, a standardized survey that illuminates not only the success of the procedure but also the various cognitive loads a novice trainee experiences throughout the simulation-based training (SBT). The NASA TLX appears to be a valuable tool to help bridge the deficiencies in MR literature as it provides insights into the various procedural workloads that are relevant to training a novice \autocite{hart2006nasa}. The NASA TLX is often used in healthcare to assess various physical and mental demands that are inherent to performing medical procedures \autocite{al2010simulation}. Within healthcare, the NASA TLX is often used for procedures performed on real patients as opposed to simulations, which is largely why the SIM TLX was developed \autocite{Harris2020}, \autocite{10.1002/bjs.10795}. The SIM TLX is used to discern between the various cognitive loads that are unique to simulation environments. The combination of both tools may be useful for assessing the efficacy of MR or augmented reality (AR) systems that aid in SBT environments.

Even fewer studies have used validated evaluation tools to assess procedures like ultrasound-guided central venous line insertion (US-CVC. Nguyen \etal \autocite{us-mr}  experiment with a real-time ultrasound display in MR, a feature that we find to be useful in our system. Rochlen \etal overlaid internal vascular anatomy on a mannequin to enhance central venous catheter insertion training \autocite{rochlen2017first}. Independent raters used a checklist to grade trainee knowledge and needle placement, however, overall performance using a procedural checklist was not measured. They additionally used a survey instrument to elicit trainee perceptions of the technology, which was positively received.

\section{Elicitation Study} \label{sec:elicitation}
\begin{figure}
    \centering
    \includegraphics[width=\columnwidth]{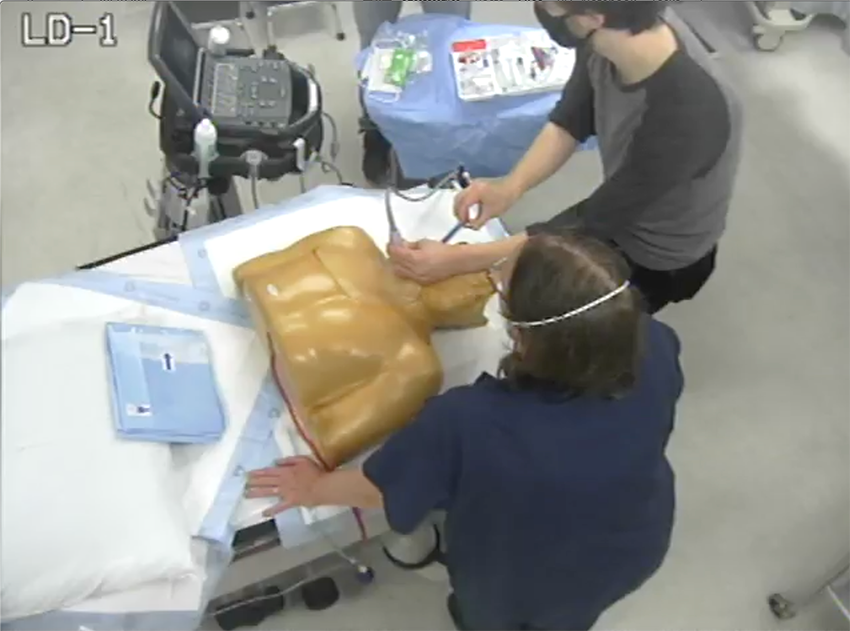}
    \caption{In-person training. The trainee and the instructor at the US-CVC simulation station.}
    \label{fig:uscvc}
\end{figure}

In order to identify the requirements of an MR system for the US-CVC procedure, we first need to analyze and understand the types of interactions and communication occurring during US-CVC procedural training. Thus, we performed a formal requirements elicitation study. The study seeks to elicit a taxonomy of nonverbal, and gestural communications between the mentor and mentee during an in-person simulator-based training of US-CVC placement and to assist in the development of the essential functional requirements of the MR training environment. We also seek to better understand how instructors adapt their practices to teach US-CVC placement to trainees with different skills and experience. An example scene showing the instructor teaching the US-CVC procedure to the trainee during the elicitation study is presented in \Cref{fig:uscvc}. 

\subsection{Participants}
We have two types of participants in our study: procedure instructors and novice trainees. We recruited senior residents, fellows, and faculty physicians with experience in US-CVC placement and medical education as instructors and medical and allied health professions students, including those on clinical rotations, as trainees. Recruitment was initiated via email and flyers posted in medical trainee team offices, lounges, and locker rooms. The study received human subject research approval from the Institutional Review Board. Altogether 10 instructors (4 male/6 female, between less than a year and 10 years of CVC teaching experience, average 34 years old) and 10 trainees (3 male/7 female, on average: 30 years old, 5 years of clinical experience), all lived and trained in the USA, have participated in the study. 

\subsection{Procedure and Apparatus} \label{sec:cvc-procedure}
Learning the US-CVC procedure consisted of two stages. In the first stage, instructors/trainees received standard US-CVC training material prior to a hands-on training session. The hands-on training was performed in 1:1 instructor/trainee pairs. These pairs were placed in individual rooms with a US machine and US-CVC simulator, following all COVID-19-related safety protocols. The training session started with an outline of the procedural steps, the preparation of the CVC kit, and the use of the US device. Key portions of the procedure, including ultrasound guided venous access, passing of a wire through the needle, and catheter placement (Seldinger technique) were completed. Correct wire placement in the target vessel was confirmed via US. The procedure was completed when the catheter was in place, and all three ports could be flushed with saline.
	
\subsection{Data Collection} \label{sec:data-collection}
Subjects were videotaped during a standard training session for US-CVC and a subsequent debriefing. 
At the end of the session, instructors and trainees were asked to complete questionnaires about their experience with the training process which were specifically related to workload and types of communications and teaching techniques used. The instruments we used include NASA TLX and SIM TLX forms. 

\subsection{Results and Discussion}

\begin{figure}
        \centering

\begin{tikzpicture}  
\begin{axis}  
[  bar width=4, 
    ybar, 
    enlargelimits=0.08,
    legend style={at={(0.5,1.1)}, 
      anchor=north, legend columns=2},     
    ylabel={SIM TLX score}, 
    ylabel style={},
    yticklabel style={rotate=90},
    ymax=15,
    symbolic x coords={Mental D.,Physical D., Temporal D., Frustration, Task Complexity, Situational Stress, Distractions, Perceptual Strain, Task Control},  
    xtick=data,  
    xticklabel style={rotate=90},
    nodes near coords align={horizontal}, 
    nodes near coords style={font=\tiny, rotate=90},
    width=\columnwidth,
    ]  
\addplot[fill={green1}] coordinates {(Mental D., 13) (Physical D., 7) (Temporal D., 1) (Frustration, 7) (Task Complexity, 13) (Situational Stress, 5) (Distractions, 0) (Perceptual Strain, 2) (Task Control, 9)}; 
\node [above,xshift=0.1cm,yshift=0.35cm, style={font=\tiny, rotate=90}] at (axis cs:  Mental D., 13) {13 (3)};
\node [above,xshift=0.1cm,yshift=0.35cm, style={font=\tiny, rotate=90}] at (axis cs:  Physical D., 7) {7 (6)};
\node [above,xshift=0.1cm,yshift=0.35cm, style={font=\tiny, rotate=90}] at (axis cs:  Temporal D., 1) {1 (1)};
\node [above,xshift=0.1cm,yshift=0.35cm, style={font=\tiny, rotate=90}] at (axis cs:  Frustration, 7) {7 (5)};
\node [above,xshift=0.1cm,yshift=0.35cm, style={font=\tiny, rotate=90}] at (axis cs:  Task Complexity, 13) {13 (5)};
\node [above,xshift=0.1cm,yshift=0.35cm, style={font=\tiny, rotate=90}] at (axis cs:  Situational Stress, 5) {5 (4)};
\node [above,xshift=0.1cm,yshift=0.35cm, style={font=\tiny, rotate=90}] at (axis cs:  Distractions, 0) {0 (0)};
\node [above,xshift=0.1cm,yshift=0.35cm, style={font=\tiny, rotate=90}] at (axis cs:  Perceptual Strain, 2) {2 (2)};
\node [above,xshift=0.1cm,yshift=0.35cm, style={font=\tiny, rotate=90}] at (axis cs:  Task Control, 9) {9 (5)};

\addplot[postaction={pattern=north east lines}, fill={green1}] coordinates {(Mental D., 8) (Physical D., 1) (Temporal D., 3) (Frustration, 4) (Task Complexity, 11) (Situational Stress, 3) (Distractions, 1) (Perceptual Strain, 3) (Task Control, 10)}; 
\node [above,xshift=0.33cm,yshift=0.35cm, style={font=\tiny, rotate=90}] at (axis cs:  Mental D., 8) {8 (4)};
\node [above,xshift=0.33cm,yshift=0.35cm, style={font=\tiny, rotate=90}] at (axis cs:  Physical D., 1) {1 (2)};
\node [above,xshift=0.33cm,yshift=0.35cm, style={font=\tiny, rotate=90}] at (axis cs:  Temporal D., 3) {3 (3)};
\node [above,xshift=0.33cm,yshift=0.35cm, style={font=\tiny, rotate=90}] at (axis cs:  Frustration, 4) {4 (6)};
\node [above,xshift=0.33cm,yshift=0.35cm, style={font=\tiny, rotate=90}] at (axis cs:  Task Complexity, 11) {11 (5)};
\node [above,xshift=0.33cm,yshift=0.35cm, style={font=\tiny, rotate=90}] at (axis cs:  Situational Stress, 3) {3 (4)};
\node [above,xshift=0.33cm,yshift=0.35cm, style={font=\tiny, rotate=90}] at (axis cs:  Distractions, 1) {1 (1)};
\node [above,xshift=0.33cm,yshift=0.35cm, style={font=\tiny, rotate=90}] at (axis cs:  Perceptual Strain, 3) {3 (5)};
\node [above,xshift=0.33cm,yshift=0.35cm, style={font=\tiny, rotate=90}] at (axis cs:  Task Control, 10) {10 (5)};

\legend{In-person Trainee, In-person Instructor}  
  
\end{axis}  
\end{tikzpicture}  
       \caption{In-person workload. We compare the mean weighted SIM TLX score per category for trainees (solid) and instructors (line pattern) during the US-CVC procedure. 
       }
       \label{fig:person-tlx}
\end{figure}
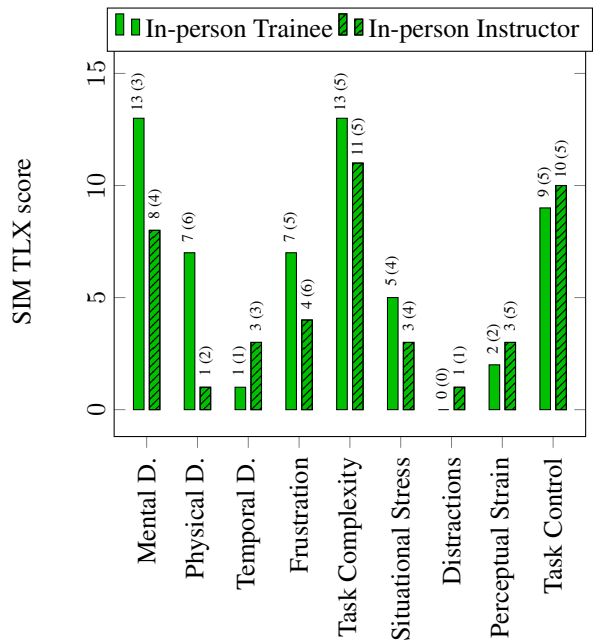

The study provided us with quantitative data on the ergonomics and various cognitive loads of procedural training, and qualitative data on the modes of interaction and forms of communication used by trainees and instructors. We measure the workload for trainees and instructors who participated in in-person training on a 0-100 scale. The total weighted workload according to the NASA TLX survey is $69 \pm 11$ and $58 \pm 10$ for the student and instructor, respectively. The SIM TLX survey results show $57 \pm 13$ and $44 \pm 14$ for the student and instructor respectively. We present the per-category SIM TLX scores in \Cref{fig:person-tlx}. 

The NASA TLX and SIM TLX were able to capture the various workloads of trainees and instructors. The survey data was analyzed using two-sample one-tailed t-tests with a significant difference of $\alpha = 0.05$. For the trainee, mental demand was the highest reported score in the NASA TLX. Trainees recorded mental workload as significantly higher relative to all other categories in the NASA TLX. According to the data, trainees also reported a higher overall workload compared to the instructors themselves. This is expected given the large gaps in experience.

In addition to the NASA TLX and SIM TLX surveys, we analyzed the videos of the training sessions to identify gestural and nonverbal communications used during training. Our goal was to identify common gestures and nonverbal cues that are used to better understand how these communications could be translated into the design of our MR medical training system. The gestures were analyzed by a medical expert who consulted with a cognitive scientist during the process. We have used the standard taxonomy of gestures with four groups: deictic, iconic, symbolic, and beat gestures \autocite{krauss_chen_gottesman_2000}.

\paragraph{Deictic Gestures} The gestures most frequently observed during procedural instruction were deictic gestures which serve to replace or emphasize words such as "that", "this", or "here". The instructor regularly employed these gestures throughout the training session to guide the trainee's attention to various locations of interest.

\begin{figure*}
    \includegraphics[width=\textwidth]{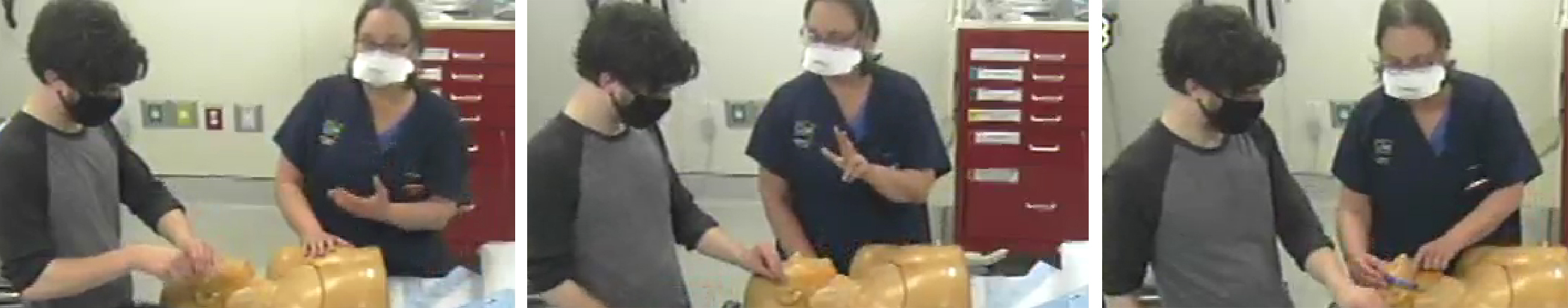}
    \caption{Gestures. Example gestures performed by the instructor: Left: Beat Gesture, Center: Iconic Gesture, Right: Deictic Gesture.}
    \label{fig:gestures}
\end{figure*}

\paragraph{Iconic Gestures} Iconic gestures were commonly observed during this procedure. These types of gestures are defined as when the instructor mimics a necessary procedural motion, perhaps accompanied with speech, to demonstrate what the trainee should do. This can be of particular utility during US-CVC where the plethora of words required to adequately describe the position, orientation, and movement of an item can be augmented by visual representation above the field.

The most complex gesture-based communication observed during this procedural training took place during the steps of the Seldinger technique, an essential part of US-CVC placement where a needle is first inserted into a blood vessel using US guidance, and then a guide wire is passed through the needle \autocite{SONG20181332}. The instructor largely relied on iconic, mimicking gestures during these steps and assisted the trainee in guiding the needle and ultrasound probe by using their hands to demonstrate the bi-manual coordinated movements of US probe and needle above the procedural field, while commenting on the US image of the needle appearing in the field of view. When providing support in techniques to hold the US probe in a stable position on the patient, the instructor referenced their own personal technique during these steps with phrases such as "this is what I would do" and "I would put my pinkie on the skin", while demonstrating a technique in which the fifth finger is stabilized on the patient's skin while the other four fingers grasp the ultrasound probe. Another integral step in the procedure is when a guide wire is threaded through the needle into the lumen of the vessel. The instructor mimicked the action of holding the wire by placing their first and second fingers together in a pinched position. While in this pinched position, the instructor would often swoop the pinched hand inward toward the mannequin to demonstrate advancing the guide wire into the lumen. 
   
During needle insertion, the operator must maintain negative pressure by drawing back on the syringe's plunger while simultaneously advancing the needle/syringe unit forward into the tissue until blood is aspirated. 
During the training session, the instructor first grasped a syringe to demonstrate these techniques, then reinforced aspiration of the syringe while the trainee performed the procedure by slightly extending and spreading fingers apart into a claw-like shape.

\paragraph{Symbolic Gestures} There were two main symbolic gestures observed in the procedural training session, one gesture symbolizing ``good'' and another symbolizing ``stop''. The gesture symbolizing ``good'' was a thumbs-up gesture performed with either one or both hands. The instructor used this gesture to recognize when the trainee had completed a task well, as a positive reinforcement technique. The other symbolic gesture observed in the session was the ``stop'' gesture which involved an open palm directed toward the trainee or the trainee's hands. This gesture was used when a trainee was doing something incorrectly or when the instructor wanted the trainee to slow down, as an adjunct to verbal instructions. The ``stop'' gesture was also used just prior to directing the trainee's attention to a separate task, along with the phrase ``let's set this up first''. 
   
\paragraph{Beat Gestures} The instructor also used several beat gestures during the session. Beat gestures are nonspecific gestures that accompany the flow of one's speech and do not have any specific meaning tied to them. The most common beat gesture observed during the training session was the instructor's slight extension and elevation of one or two hands in front of them with palms facing the ceiling. The instructor used this gesture throughout the training session as they spoke with the trainee although the gesture did not appear to be associated with any particular training points. 

\section{System Requirements} \label{sec:sys-requirements}
Based on the results of the elicitation study, we established a set of high-level system requirements for our MR volumetric communication system as well as five specific functional requirements:
\begin{enumerate*}
    \item Spatial information,
    \item Hands-free communication
    \item Hand tracking,
    \item Virtual objects, and
    \item Voice communication.
\end{enumerate*}

At the highest level of abstraction, we identified a spatial context as an important component for the instructor's ability to guide a trainee. Thus, we decided that the remote expert needs to be presented with a high-fidelity volumetric view of the local procedure. The scene needs to be presented in a way that resembles being physically co-located, because, as we observed in the study, there is frequent gestural and non-verbal communication between the instructor and the trainee.

The trainee needs both hands to perform US-CVC procedure. Additionally, sterility has to be guaranteed throughout the procedure. Thus, hands-free communication is required.

The local operator needs to receive visual guidance from the remote expert. The visual guidance can be given in the form of gestures. This can be achieved by implementing a MR virtual hand, which visualizes the representation of animated remote expert's hands in the local operator's MR space.  

In addition to gestural guidance, it may be beneficial to provide the remote expert the ability to use virtual objects to demonstrate procedural steps. Virtual objects will allow for a more intuitive communication. Object demonstrations should allow the expert to switch the focus from verbal instructions to visual instructions during parts of the procedure, \eg US probe and needle guidance.  
Besides visual telepresence, the system must provide two-way real-time audio communication.

Finally, based on the NASA TLX results showing significant difference in mental load between the trainee and the instructor, we decided to take an asymmetric approach to the use of MR in our system. In order to limit the additional mental load, resulting from using the MR system, we limit the trainee’s augmented view to the virtual objects and the expert's hand presentation only. The remote expert's view, whose mental load in our elicitation study was determined to be significantly lower than those of the trainee, is entirely depended on the MR interaction.

\begin{figure*}
    \centering
    \includegraphics[width=\textwidth]{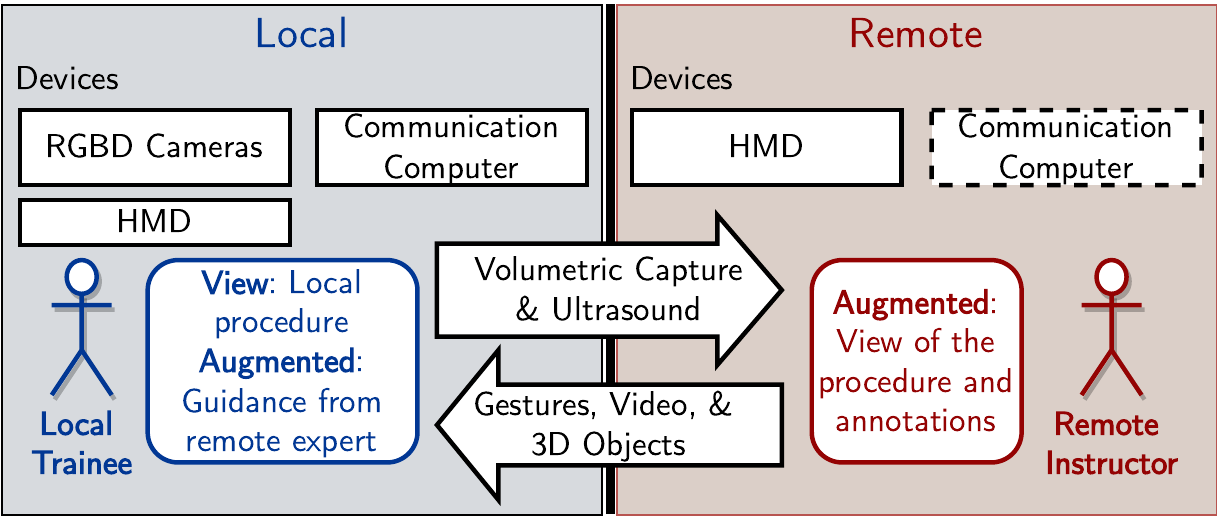}
    \caption{System diagram. We illustrate the two main actors, the local trainee (blue) and the remote instructor (red) and their views on the scene. The information transfer between the actors is indicated by arrows.}
    \label{fig:sys-diagram}
\end{figure*}

\section{Mixed Reality System}\label{sec:mr-system}

The MR system enables two-way real-time volumetric-based communication.
The two parties that our system is designed for are the remote expert and the local operator of the US-CVC procedure. We implement a one-to-one connection between them. The design of the system supports one-to-many connections such that a single expert is able to train multiple trainees at the same time.
A more detailed explanation of the system implementation can be found in \autocite{rebolISMAR}.

\subsection{System Components}
Based on the requirements identified in \Cref{sec:sys-requirements}, we design the following system components:
\begin{enumerate}
    \item RGBD cameras capture the local scene. A volumetric view is generated from the RGBD capture to provide the remote expert with spatial information and to allow for volumetric communication. 
    \item Head-mounted displays (HMDs) are used to allow for hands-free MR communication.
    \item HMDs track the user's hands to support gestural communication. 
    \item US-CVC specific objects, as well as abstract objects, are implemented. 
    \item HMDs with microphones and speakers are used for voice communication. 
\end{enumerate}

\subsection{Devices} 
The Microsoft Hololens 2 HMD was used because this MR headset supports volumetric render, head tracking, and hand tracking for visual communication. Moreover, it has a built-in microphone and speakers for voice communication. We use Microsoft Azure Kinect RGBD cameras to capture the local scene. The Azure Kinect time-of-flight (TOF) sensor's typical systematic error is below 11 mm + 0.1\% of distance which should be accurate enough to teach the US-CVC procedure. In addition to visual communication devices, we use network devices and computers to establish the connection.

\subsection{Views and Interaction} 
We illustrate the system including actors, devices, and communication flows in \Cref{fig:sys-diagram}.
The remote instructor has two options to guide the local trainee: 
\begin{itemize}
    \item The instructor can use gestures, which are captured by the Hololens 2 camera system.
    \item The instructor can use virtual objects to give directions. 
\end{itemize}

The scene is captured by RGBD cameras at the local site. The RGBD cameras are connected to a computer that processes the captured data before it is forwarded to the remote instructor. The computer is equipped with an Intel i7 10th Generation CPU and an Nvidia Geforce GTX 1080 Ti GPU to allow for real-time processing.

The local trainee wears an MR HMD through which they receive audio and visual guidance from the remote instructor. Abstract objects such as cuboids and cylinders as well as realistic renderings of medical tools from the remote instructor are augmented onto the local scene through the trainee's Hololens. The gestures captured from the remote expert are visualized using a virtual 3D hand model which allows the local operator to receive directions. For visual interaction in the local trainee's physical environment, we align the RGBD cameras, HMDs, and the physical workspace.
\subsection{Data Communication} \label{sec:data-communication}
Our system consists of the following communication nodes which transfer data between each other: the camera computer, the remote instructor's HMD, the local trainee's HMD, and the remote computer. Because of the high bandwidth needed for real-time volumetric communication, we propose a strategy that supports low-latency and low-bandwidth volumetric communication. We send depth and RGB color frames separately over the network to utilize existing compression codecs. For RGB video, the MR - WebRTC 
built-in VP9 \autocite{vp9} encoding is used. The encoding allows us to send 1920 x 1080 pixel color at 30 frames per second (FPS) with a target bit rate of 1.9 Mbps. The depth images are sent keyframe encoded. The whole depth image is sent once every second. In between, only depth pixels that changed between frames are sent. We compress the depth image using the Deflate algorithm \autocite{DEFLATE} based on LZ77 \autocite{lz77} and Huffman coding \autocite{huffman-coding}. 
This ensures bandwidth consumption below 1.9 Mbps assuming a scene change below 10 \% between frames and a depth resolution of 320 x 288 pixels at 30 FPS. We observe that the depth scene change is mostly below 10 \% during the US-CVC procedure when using a static camera position. 
The temporal coherence computation on the GPU rendering takes about 5ms per frame on a laptop. Thus, latency is dominated by the network connection. 
The devices to communicate at 30 FPS on a local network with a Wi-Fi connection. The latency is unnoticeable on a local network.

\subsection{Software Tools}
We build our MR application for the Hololens 2 HMDs based upon the Mixed Reality Toolkit \autocite{mrtk}. The toolkit allows us to support standardized MR user interaction. On the remote instructor's Hololens, we used Mixed Reality Toolkit's hand tracking  
to predict the hand position. The Hololens 2 head-tracking 
allows for stable alignment between the nodes. 
We implemented the MR WebRTC client to manage the communication between the four nodes. The video data including volumetric RGB, webcam feed, and US feed, is sent over WebRTC channels. We use separate web-socket-based data channels to forward depth, position, rotation, and annotation data. 

\section{Evaluation} \label{sec:evaluation}
We evaluated the proposed mixed reality (MR) communication system in a study in which we compare it against video communication. We selected 20 medical students (5 male/14 female, on average: 27 years old, 2 years clinical experience, no AR/VR experience) and 5 medical experts (4 male/1 female, 1+ years of CVC teaching experience, average 43 years old), all lived and trained in the USA. They were randomly assigned into 10 MR teaching and 10 video teaching pairs. In the study, the medical experts taught the ultrasound-guided CVC procedure to the students. The US-CVC training was performed on US simulation mannequins. We used similar camera position setups for video and MR to ensure a fair comparison. Although the MR system is designed for remote communication, we conducted the study in a lab setting on a local network. The evaluation steps and the data collection were the same as in the elicitation study (\Cref{sec:cvc-procedure}, \Cref{sec:data-collection}). Yet, we switched to a less detailed US-CVC training method because of time constraints. Thus, elicitation results should not be compared directly. After an introductory video and an initial CVC kit walk-through, the remote training of the procedure started. 

\begin{table}[]
	\setlength\extrarowheight{2pt} 
	\setlength{\tabcolsep}{7pt}
	\footnotesize
	\centering
	\begin{tabular}{r||c|c||c|c}
		&\multicolumn{2}{c||}{\textbf{NASA TLX}} & \multicolumn{2}{c}{\textbf{SIM TLX}}  \\ 
		&Video & MR & Video & MR\\ \hline \hline
		\textbf{Trainee} & $64 \pm 10$ & $65 \pm 9$ & $49 \pm 18$ & $49 \pm 20$\\ \hline
		\textbf{Instructor} & $63 \pm 13$ & $67 \pm 15$ & $48 \pm 18$ & $54 \pm 16$\\ \hline \hline
	\end{tabular}
	\caption{NASA TLX and SIM TLX workload for trainees and instructors during US-CVC training using videoconferencing and MR technology.
	}\label{tab:results}
\end{table}	

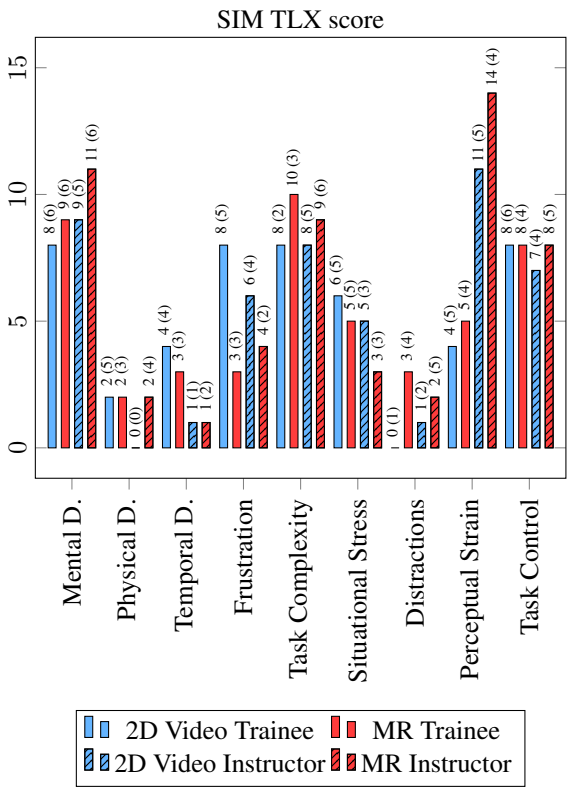
\begin{figure}[!ht]
    \centering

\begin{tikzpicture}  
\begin{axis}  
[   bar width=3, 
    title={SIM TLX score},
    title style={at={(0.5,1)}, anchor=north, yshift=8.0},
    ybar, 
    enlargelimits=0.08,
    legend style={at={(0.5,-0.7)}, anchor=south, legend columns=2},  
    ylabel style={},
    yticklabel style={rotate=90},
    ymax=15,
    symbolic x coords={Mental D., Physical D., Temporal D., Frustration, Task Complexity, Situational Stress, Distractions, Perceptual Strain, Task Control},  
    xtick=data,
    xtick pos=left,
    xticklabel style={rotate=90},
    nodes near coords align={horizontal}, 
    nodes near coords style={font=\tiny, rotate=90},
    width=1.1\columnwidth,
    ]  
\addplot[fill={blue1}] coordinates {(Mental D., 8) (Physical D., 2) (Temporal D., 4) (Frustration, 8) (Task Complexity, 8) (Situational Stress, 6) (Distractions, 0) (Perceptual Strain, 4) (Task Control, 8)}; 
\node [above,xshift=-0.07cm,yshift=0.35cm, style={font=\tiny, rotate=90}] at (axis cs:  Mental D., 8) {8 (6)};
\node [above,xshift=-0.07cm,yshift=0.35cm, style={font=\tiny, rotate=90}] at (axis cs:  Physical D., 2) {2 (5)};
\node [above,xshift=-0.07cm,yshift=0.35cm, style={font=\tiny, rotate=90}] at (axis cs:  Temporal D., 4) {4 (4)};
\node [above,xshift=-0.07cm,yshift=0.35cm, style={font=\tiny, rotate=90}] at (axis cs:  Frustration, 8) {8 (5)};
\node [above,xshift=-0.07cm,yshift=0.35cm, style={font=\tiny, rotate=90}] at (axis cs:  Task Complexity, 8) {8 (2)};
\node [above,xshift=-0.07cm,yshift=0.35cm, style={font=\tiny, rotate=90}] at (axis cs:  Situational Stress, 6) {6 (5)};
\node [above,xshift=-0.07cm,yshift=0.35cm, style={font=\tiny, rotate=90}] at (axis cs:  Distractions, 0) {0 (1)};
\node [above,xshift=-0.07cm,yshift=0.35cm, style={font=\tiny, rotate=90}] at (axis cs:  Perceptual Strain, 4) {4 (5)};
\node [above,xshift=-0.07cm,yshift=0.35cm, style={font=\tiny, rotate=90}] at (axis cs:  Task Control, 8) {8 (6)};

\addplot[fill={red1}] coordinates {(Mental D., 9) (Physical D., 2) (Temporal D., 3) (Frustration, 3) (Task Complexity, 10) (Situational Stress, 5) (Distractions, 3) (Perceptual Strain, 5) (Task Control, 8)}; 
\node [above,xshift=0.12cm,yshift=0.35cm, style={font=\tiny, rotate=90}] at (axis cs:  Mental D., 9) {9 (6)};
\node [above,xshift=0.12cm,yshift=0.35cm, style={font=\tiny, rotate=90}] at (axis cs:  Physical D., 2) {2 (3)};
\node [above,xshift=0.12cm,yshift=0.35cm, style={font=\tiny, rotate=90}] at (axis cs:  Temporal D., 3) {3 (3)};
\node [above,xshift=0.12cm,yshift=0.35cm, style={font=\tiny, rotate=90}] at (axis cs:  Frustration, 3) {3 (3)};
\node [above,xshift=0.12cm,yshift=0.35cm, style={font=\tiny, rotate=90}] at (axis cs:  Task Complexity, 10) {10 (3)};
\node [above,xshift=0.12cm,yshift=0.35cm, style={font=\tiny, rotate=90}] at (axis cs:  Situational Stress, 5) {5 (5)};
\node [above,xshift=0.12cm,yshift=0.35cm, style={font=\tiny, rotate=90}] at (axis cs:  Distractions, 3) {3 (4)};
\node [above,xshift=0.12cm,yshift=0.35cm, style={font=\tiny, rotate=90}] at (axis cs:  Perceptual Strain, 5) {5 (4)};
\node [above,xshift=0.12cm,yshift=0.35cm, style={font=\tiny, rotate=90}] at (axis cs:  Task Control, 8) {8 (4)};

\addplot[postaction={pattern=north east lines}, fill={blue1}] coordinates {(Mental D., 9) (Physical D., 0) (Temporal D., 1) (Frustration, 6) (Task Complexity, 8) (Situational Stress, 5) (Distractions, 1) (Perceptual Strain, 11) (Task Control, 7)}; 
\node [above,xshift=0.31cm,yshift=0.35cm, style={font=\tiny, rotate=90}] at (axis cs:  Mental D., 9) {9 (5)};
\node [above,xshift=0.31cm,yshift=0.35cm, style={font=\tiny, rotate=90}] at (axis cs:  Physical D., 0) {0 (0)};
\node [above,xshift=0.31cm,yshift=0.35cm, style={font=\tiny, rotate=90}] at (axis cs:  Temporal D., 1) {1 (1)};
\node [above,xshift=0.31cm,yshift=0.35cm, style={font=\tiny, rotate=90}] at (axis cs:  Frustration, 6) {6 (4)};
\node [above,xshift=0.31cm,yshift=0.35cm, style={font=\tiny, rotate=90}] at (axis cs:  Task Complexity, 8) {8 (5)};
\node [above,xshift=0.31cm,yshift=0.35cm, style={font=\tiny, rotate=90}] at (axis cs:  Situational Stress, 5) {5 (3)};
\node [above,xshift=0.31cm,yshift=0.35cm, style={font=\tiny, rotate=90}] at (axis cs:  Distractions, 1) {1 (2)};
\node [above,xshift=0.31cm,yshift=0.35cm, style={font=\tiny, rotate=90}] at (axis cs:  Perceptual Strain, 11) {11 (5)};
\node [above,xshift=0.31cm,yshift=0.35cm, style={font=\tiny, rotate=90}] at (axis cs:  Task Control, 7) {7 (4)};

\addplot[ postaction={pattern=north east lines}, fill={red1}] coordinates {(Mental D., 11) (Physical D., 2) (Temporal D., 1) (Frustration, 4) (Task Complexity, 9) (Situational Stress, 3) (Distractions, 2) (Perceptual Strain, 14) (Task Control, 8)}; 
\node [above,xshift=0.49cm,yshift=0.35cm, style={font=\tiny, rotate=90}] at (axis cs:  Mental D., 11) {11 (6)};
\node [above,xshift=0.49cm,yshift=0.35cm, style={font=\tiny, rotate=90}] at (axis cs:  Physical D., 2) {2 (4)};
\node [above,xshift=0.49cm,yshift=0.35cm, style={font=\tiny, rotate=90}] at (axis cs:  Temporal D., 1) {1 (2)};
\node [above,xshift=0.49cm,yshift=0.35cm, style={font=\tiny, rotate=90}] at (axis cs:  Frustration, 4) {4 (2)};
\node [above,xshift=0.49cm,yshift=0.35cm, style={font=\tiny, rotate=90}] at (axis cs:  Task Complexity, 9) {9 (6)};
\node [above,xshift=0.49cm,yshift=0.35cm, style={font=\tiny, rotate=90}] at (axis cs:  Situational Stress, 3) {3 (3)};
\node [above,xshift=0.49cm,yshift=0.35cm, style={font=\tiny, rotate=90}] at (axis cs:  Distractions, 2) {2 (5)};
\node [above,xshift=0.49cm,yshift=0.35cm, style={font=\tiny, rotate=90}] at (axis cs:  Perceptual Strain, 14) {14 (4)};
\node [above,xshift=0.49cm,yshift=0.35cm, style={font=\tiny, rotate=90}] at (axis cs:  Task Control, 8) {8 (5)};

\legend{2D Video Trainee, MR Trainee, 2D Video Instructor, MR Instructor}  
  
\end{axis}  
\end{tikzpicture}  
       \caption{Video vs MR workload. We compare the mean weighted SIM TLX score for trainee and instructor using video (red) and MR (blue). Score and workload category are shown on the y and x-axis, resp. The instructor bars are filled with a line pattern.}
       \label{fig:video-mr}
\end{figure}

\subsection{Results and Discussion}
The total weighted subjective workload results of the NASA TLX and the SIM TLX survey are shown in \Cref{tab:results}. A two-sample Student's t-test does not show a significant difference for $\alpha = 0.05$ between the video and MR training for the instructor and the trainee group. 

When examining the per-category workload results, we find no significant difference in the NASA TLX survey but two categories of the SIM TLX survey show a significant difference.
A two-sample one-tailed t-test in both directions shows significantly lower frustration (p=0.01) but higher distraction (p=0.02) for trainees when using MR. We assume a higher frustration workload for the trainee using video came from the fact that complex parts of the procedure were harder to understand because of the limitations such as missing spatial information of single-view 2D video when using visual explanations. Higher distraction with MR could be due to the higher amount of visual information the trainee is presented with. 
We present the per-category SIM TLX results in \Cref{fig:video-mr}.

The main finding is that workload does not increase significantly when using MR for US-CVC training, compared to video communication. 
We conclude that the MR system we designed does not add additional workload for the trainee and the instructor during complex medical procedural training compared to video despite the novelty of the technology. We see opportunities for better performance using MR compared to video communication because of spatial information, virtual tools, and gestural communication.   

We have shown that MR can be used for complex procedural training when designed strategically in consultation with domain experts. The participants of our study received instruction on Microsoft Hololens 2 HMD adjustment and use, but did not receive MR training or practice time before using the technology to complete the procedure. We argue that MR systems can be used for various remote training situations \autocite{cpr-work-in-progress} such as machine operations and repair, because of low technology training requirements and the successful use in the complex medical US-CVC procedure in which precise communication is required. However, we believe that an iterative design and consultation with domain experts is needed to increase the chances of successful MR technology usage.

Moreover, we argue that the technology could be used in an emergency setting due to many similarities between our study's simulation and an emergency situation. We identify the potential of lower workload when using MR compared to video communication for experienced users. There could be a learning curve when teaching through MR because of various opportunities for teaching such as gestures, virtual objects, and annotations. We observe that the instructors adapt their teaching style after teaching a second MR session.  
In future work, we want to examine how MR teaching changes over time when the same instructors teach multiple times. 

\section{Conclusion}
We presented the design process of an MR communication system for medical procedures. The design requirements for the US-CVC procedure were analyzed in an elicitation study. After design analysis, we implemented the system focusing on lightweight architecture meaning: affordable consumer devices, easy and fast setup, low network bandwidth, and low computational requirements. 

We evaluated the final system in a study in which we compare video against MR on simulated US-CVC placement training. The subjects do not receive any specific MR training before the procedure other than a short briefing on the functionality. The self-reported workload results indicate that workload load does not increase significantly when using MR compared to videoconferencing. Thus, we conclude that MR systems can be used intuitively for complex training procedures when designed strategically with domain experts. 
In future experiments, we plan to analyze how MR training with the proposed system changes over time as MR instructors become more experienced.

\section{Acknowledgments}
The work is supported by National Science Foundation grant no. 2026505 and 2026568. 
The authors wish to thank David Li, Rohan Patil, Conor Schulte, Erin Horan, Safinaz Alshiakh, Yasser Ajabnoor, Jesse Schwartz, Becky Lake, Scott Schechtman, and Carine Cristina Goncalves Galvao for their help.










\printbibliography

\end{document}